# Proposed Universal Relationships Describing Electron Phonon Mediated Superconductivity and Accommodation of the Cuprates and Pnictides

D.A. Nepela
4505 7th St. SE, Puyallup, WA 98374

**Abstract**

A universal and self-consistent set of equations is developed utilizing the principle empirical parameters of Superconductivity i.e. $T_c$, $\theta_D$, $\Delta_{c(o)}$, $\lambda$, which are the coherent condensation temperature, the Debye temperature, the coherent condensation energy gap at $T_c= 0°K$ and the electron phonon coupling constant respectively. Empirical data from both crystalline elements and amorphous compounds is shown to produce the same self consistent relationships which are $T_c= \theta_D/2 \exp(-2/\lambda)$, $\Delta_{c(o)}=\hbar\omega_D \exp(-2/\lambda)$ and the ratio of $\Delta_{c(o)}/T_c= 2.0$ i.e. $\Delta_{c(o)}= 2k_B T_c$ and is found to be constant for all superconductors. We also find that $\lambda = -2/\ln(E_{(o)}/T_c)$ where $E_{(o)} = \theta_D/2$ and is the zero point energy of a quantum mechanical oscillator. These relationships are derived exclusively from electron phonon mediated superconductor data and are also shown to encompass cuprate superconductors with ease. Our self-consistent relationships lead to other important relationships which are that the maximum possible $\lambda$ for electron phonon mediated superconductivity is $\lambda_{max}= 2.89$. Also by use of first principle calculations of the shift in chemical potential $\Delta\mu$ with doping for both hole and electron doped cuprates show that $\Delta\mu$ at peak $T_c$ is identical to $\Delta_{c(o)}$ and we can, from chemical shift data, define the entire dome shaped $T_c$ vs. x for $La_{2-x}Sr_xCuO_4$ and show that the peak occurs at x = 0.1675 Sr concentration which is in excellent agreement with eperimental data. The use of $\eta = 2T_c/E_o$, = efficiency of phonon utilization of the source energy $E_o$. When utilized to compare hole vs. electron doped cuprates clearly incicates that hole doped cuprates are 1.76 times as efficient in utilizing the available phonon energy from $E_o$,

We further develop arguments from which we conclude the pseudogap associated with cuprates has no role in their superconductivity. Similar arguments are given that the mechanism of superconductivity remains electron phonon mediated with no need for any energy scale changes from qualified electron phonon mediated superconductivity. Additionally we have found with limited information on the new pnictides that they conform to our universal modified BCS equations and are thus electron-phonon mediated.

## Introduction

From the onset of BCS[1] Theory Bardeen, Cooper, and Schrieffer in 1957 through its validation in 1976 (Nobel Prize) there existed and remained residual difficulties in being able to reliably calculate (estimate) the superconducting critical temperature $T_c$. that was expressed by BCS as:

Eq.1: $T_c = 1.13\theta_D \exp[-1/ NoV]$

wherein No is the density of states at $E_f$ (FermiEnergy) and V is some average matrix element of the electron attraction which corresponds to the attraction between electrons in the BCS theory.

BCS in their 1957 publication also indicated the equation for the energy gap $\Delta_o$ which was defined as the coherent condensation energy gap defined as:

Eq. 2: $\Delta_o = 2\hbar\omega_D e^{-1/NoV}$

and also found the ratio of $\Delta_o/T_c$ to be a constant equal to 1.76 which was expressed as:

Eq. 3: $\Delta_o = 1.76 k_B T_c$

After the publication of BCS theory, a number of refinements of the initial theory developed, these were Migdal's[2] theory of electron phonon effects in the normal state and the Eliashberg[3] theory which generalizes BCS' theory to incorporate Migdal[2] theory in the limit and to provide a more rigorous basis for $T_c$. Additionally Morel and Anderson[4] introduced $\mu^*$ to account for the effects of electron-electron coulomb repulsion wherein $\lambda-\mu^*$ replaces NoV of the BCS equation[1] and we have:
Eq. 4: $T_c = 1.13\theta_D \exp[-1/\lambda-\mu^*]$

The BCS Theory of Superconductivity in it's essentials, consists of the pairing of electrons via an interaction with lattice vibrations (phonons) which is characterized by $\lambda$ the electron-phonon coupling constant (related to the average separation of the pairs when formed and, the binding energy between the components of the pair. This pair state as Cooper pairs which ultimately undergoes at $T_c$ a condensation and thus becomes the current carrying element of resistanceless current flow, i.e. superconductivity.

The BCS Theory as formulated derives its energy scale for the electron-phonon interaction by <u>assuming</u> that the appropriate energy scale for superconductivity adjacent $E_f$ is $\leq \hbar\omega_D$ i.e. the Debye energy thus the phonon connection.

The BCS Theory was formulated as a weak coupling theory i.e. appropriate for $\lambda<<1.0$.

Early on it became apparent that the theory did not do well for stronger coupling elements such as Hg and Pb and that in fact $\Delta_o/T_c \neq 1.76$ but increased to above 2.2 for a number of superconductive compounds as we indicate in Appendix 1. The problem was further exacerbated in 1986 by the discovery of the cuprates by Bednorz and Mueller[5] and, their later follow ons with $T_c$'s >100°K and $\Delta_o/T_c \geq 2.5$. Since the discovery of cuprates many new proposals and theories have emerged attempting to understand and to incorporate the cuprates into such new theories. All attempts of this kind have failed to gain general acceptance. Thus, the aforesaid inadequacy of the BCS Theory to accommodate all experimental data including cuprates remains a problem seeking a solution.

## Development of a Self Consistent Set of Expressions for Superconductivity Utilizing Experimental Data

A complex system such as superconductivity provides a base of experimental data that has been accumulated over the last 50 years from the discovery of tunneling by Giaver[6] and its ability to extract relevant superconductor parameters independently i.e. the gap energy $\Delta_o$ which we have redefined as $\Delta_{E(o)}$ consisting of $\Delta_{c(o)}$ (coherent condensation gap and an additional energy term) which we will subsequently discuss. Also, the development of McMillians[7] inversion procedures for the tunneling data from which $\lambda$ and $\mu^*$, the so called pseudopotential, could be extracted. Unfortunately for most cuprates, the tunneling $\Delta_o$ is difficult to obtain for cuprates due to inadequate sample quality and sample interface uncertainties. It has been reliably obtained for several cuprates only.

Our approach is to use this rich source of reliable and reproducible data on various superconductors <u>which inherently interrelate the complex interrelationships of the parameters important to superconductivity</u>. The remaining question then is how one gives voice to the experimental data such that the proper relationships to the parameters of interest are properly disposed from one superconductor to any other. The approach was to incorporate in our use of the experimental data, those elements or parameters we believed to be true or reasonable. For example, Frohlich[8] in 1950 before the BCS Theory of 1957 found that the energy attendant to superconductivity was proportional to $\exp(-1/\lambda)$ and BCS found a proportionality of the same type. i.e. proportional to $\exp(-1/NoV)$ where NoV is essentially equal to $\lambda$. Accordingly we used this form in our work to determine the interrelationships between the relevant parameters of superconductivity.

The specifics of this form are open to modification by interaction with experimental data. Both McMillian[7] and Allen-Dynes[9] found $\mu^*$ (coloumb pseudopotential) to be generally ~ 0.10 for most superconductors we therefore held this parameter constant to work with the experimental data sets we used. We also found that $\Delta_o/T_c \neq 1.76$ as was found by BCS but that from approximately 30 well measured and reproducible values of $\Delta_{E(o)}$ and $T_c$ we determined the average value to be ~2.0. See Appendix 1.

**Physically Rationalized Relationship of Superconductive Parameters Required for Extraction of Analytic Expressions Describing Superconductivity via Experimental Data Input**

We now proceed to provide the physical arguments for our choice of variables and their form yielding a clear possibility of arriving at a unique and physically relevant relationship between the variables. We begin with the exponential expression relating $T_c$ to $\lambda$, and $\Delta_o$ which we elect to retain since both Froelich[8] and BCS[1] found expressions of identical form. Froelich found that $T_c \propto \exp(-1/\lambda)$ whereas BCS found $T_c \propto \exp(-1/NoV)$ where $\lambda \approx NoV$, and later it was found that the use of the pseudopotential to be useful as well thus resulting in $\lambda \propto \exp(-1/\lambda-\mu^*)$ which form we will use to analyze experimental $\lambda$ inputs from the McMillan[7] equations and data sets. The power of this exponential will be decided by the experimental inputs of $\lambda$'s against our counterpart variable choices. The counterpart part of the problem involves either $\Delta_{c(o)}$, the coherent condensation gap, $T_c$ and $\theta_D$ (Debye temperature) which provides the <u>phonon connection</u> for our analysis. Our choice of $\lambda$ was based on the fact that this parameter has the property that as $\lambda \to \infty$, the strength of electron phonon coupling increases to maximum values and accordingly $\lambda^{-1} \to 0$ as the coupling strength goes to a maximum as well.

Our counterpart variables must also be constructed to preserve two important limits as well. We elect the ratio of $\Delta_{c(o)}/\theta_D$ or its equivalent $2T_c/\theta_D$ since $\Delta_{c(o)} = 2T_c$. By choosing this ratio we fulfill both limiting conditions which are:

1. $2T_c/\theta_D \leq 1.0$ is a clear statement that the variable phonon energy expressed as $\theta_D$ cannot exceed unity. We will, in a later section demonstrate $(2T_c/\theta_D)^{-1} \leq 1.0$ which is also related to the efficiency with which pair formation via electron-phonon interactions occurs.

2. Equally important the $\ln(\theta_D/2T_c) \to 0$ as $\lambda^{-1} \to 0$ and by utilizing the ln of their defined ratio we <u>retain the exponential relationship with $\lambda$.</u>

By careful choice of our variable as above we have constructed a scenario wherein we have the major parameters of superconductivity disposed in such a way as to have them both simultaneously approach zero as the strength of electron phonon coupling increases and our expectation when we use the McMillan data set of $\lambda$'s for various elements at a constant, $\mu^* = 0.10$ that the plot of $\lambda^{-1}$ vs. $\ln(\theta_D/2T_c)$ will have a zero, zero intercept and from the slope and the intercept we will be able to extract the proper analytic expression for $T_c$ as determined from the experimental $\lambda_{McMillian}$ inputs. Thus, yeilding a linear analytic relationship between $T_c$ and the other relevant parameters of superconductivity expressed as:

Eq. 5:  $T_c = \theta_D/2 \exp(-2/\lambda)$; and

Eq. 5a represents $\lambda = -2/\ln(E_o/T_c)$

The (zero,zero) intercept <u>independently</u> verifies that $2T_c = \Delta_{c(o)}$ via experimental inputs, thus validating our initial assumption from the data of Appendix 1 which yielded a relationship of $\Delta = 2T_c$. We now find that

Eq. 5b:  $\Delta_{c(o)} = 2E_o \exp(-2/\lambda)$

and their ratio is:

Eq. 6:  $\Delta_{c(o)}/T_c = \dfrac{2E_o \exp(-2/\lambda)}{E_o \exp(-2/\lambda)} = 2.0$  or alternately

$\Delta_{c(o)} = 2k_B T_c$

We have just validated through experimental data that $\Delta_{c(o)}$ the coherent condensation energy gap is equal to $2k_B T_c$. This same relationship has been theoretically demonstrated by X. H. Zheng[9a] where he proves that $\Delta = 2k_B T_c$ where Zheng adopts $\Delta$ as the BCS coherent condensation energy gap, ie Zheng's $\Delta$ is equal to our $\Delta_{c(o)}$ thus, providing a theoretical basis for our finding of the same value by utilization of experimental data.

Figure 1 illustrates that this analytic rela-tionship is $T_c = \theta_D/2\exp(-2/\lambda)$ thus the experimental data $\lambda_{McMillian}$ input has been successfully used to establish the magnitude of the exponential term

*Figure 1*

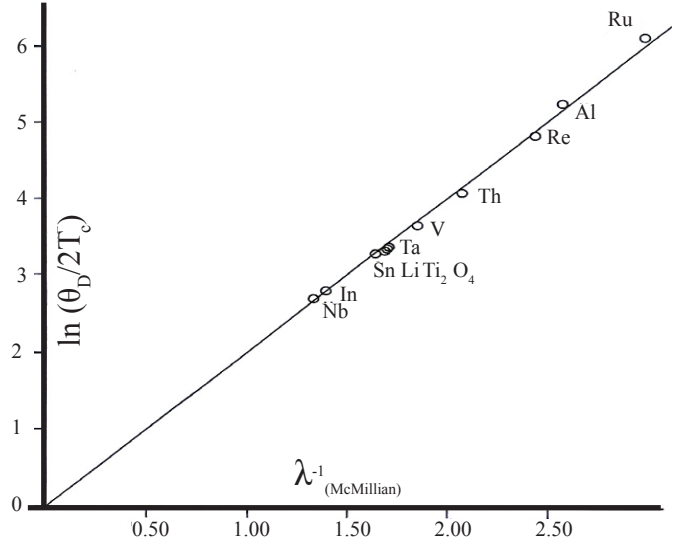

*Figure 1 illustrates the response of the ln of the dimensionless ratio $\theta_D/2T_c$ to the $\lambda_{McMillian}$ value at a constant $\mu^* = 0.10$ for a number of crystalline superconductive elements.*

and the prefactor thereto as well. We find the BCS equation to differ on both counts the prefactor is not $1.13\,\theta_D$ but $\theta_D/2 = E_o$ and in energy terms is $\frac{1}{2}\hbar\omega_D$ and the exponential we find is $T_c \propto \exp(-2/\lambda)$ whereas the BCS value is essentially $T_c \propto \exp(-1/\lambda)$. These differences are not surprising since the BCS expressions were derived for weak coupling where $\lambda < 1.0$ and thus were at a minimum necessarily limited in their scope and/or accuracy.

$$\text{We note that } 2T_c/E_o = \frac{\Delta_{c(o)}}{E_o} = 2\exp(-2/\lambda)$$

wherein $2T_c/E_o$ represents the fraction of carriers undergoing EPI (electron-phonon interaction) whereas $\exp(-2/\lambda)$ represents the fraction of Cooper pairs formed for a specific material. Quantifying these relationships for say, $YBa_2Cu_3O_7$ which has at optimal doping at $T_c \sim 92°K$. The value of $2T_c/E_{o(Y123)} = 0.842$ which means that the fraction of free carriers in $Y_{123}$ undergoing electron phonon interaction (EPI) is 0.842 whereas $\exp(-2/\lambda) = 0.421$ which means that the fraction of carriers having undergone EPI ½ of these result in Cooper pairs.

*Table I*

| Elements Compound | $\theta_D$,°K | $E_o$,°K | $\lambda$ | $T_c$,°K | $\Delta_{E(o)}$ (mev) | $\Delta_{c(o)}$ (mev) | $\Delta_{E(o)}/\Delta_{c(o)}$ | $2T_c/E_o$ |
|---|---|---|---|---|---|---|---|---|
| $YBa_2Cu_3O_7$ | 437 | 218.5 | 2.312 | 92 | 20.0 | 15.86 | 1.261 | 0.842 |
| $La_{1.85}Sr_{.15}CuO_4$ | 368 | 184 | 1.267 | 38 | 7.5 | 6.55 | 1.145 | 0.413 |
| $Tl_2Ba_2CuO_6$ | 363 | 181.5 | 2.85 | 90 | 20.5 | 15.52 | 1.32 | 0.99 |
| $LiTi_2O_4$ | 657 | 328.5 | 0.595 | 11.4 | 18 | 1.97 | 0.95 | 0.069 |
| $MgB_2$ | 915 | 451.5 | 0.816 | 39.5 | 7.1 | 6.83 | 1.04 | 0.173 |
| $HfV_2$ | 177 | 88.5 | 0.832 | 8.0 | 1.45 | 1.38 | 1.05 | 0.18 |
| $PbMo_6S_8$ | 95 | 47.5 | 1.45 | 12.0 | 2.4 | 2.07 | 1.16 | 0.505 |
| Nb | 277 | 138.5 | 0.739 | 9.25 | 1.55 | 1.59 | 0.972 | 0.134 |
| Pb | 105 | 5.25 | 1.007 | 7.2 | 1.33 | 1.24 | 1.071 | 0.274 |
| Ta | 258 | 126 | 0.595 | 4.49 | 0.72 | 0.774 | 0.93 | 0.071 |
| $Nb_3Sn$ | 228 | 114 | 1.067 | 17.5 | 3.39 | 3.16 | 1.074 | 0.307 |
| $Nb_3Al$ | 292 | 146 | 0.927 | 16.4 | 3.04 | 2.83 | 1.075 | 0.255 |
| $Nb_3Ge$ | 290 | 145 | 1.086 | 23 | 4.16 | 3.97 | 1.049 | 0.317 |
| $V_3Si$ | 465 | 232.5 | 0.765 | 17 | 2.78 | 2.93 | 0.949 | 0.146 |
| In | 112 | 56 | 0.715 | 3.14 | 0.541 | 0.589 | 0.92 | 0.122 |
| Sn | 200 | 100 | 0.608 | 3.72 | 0.593 | 0.642 | 0.925 | 0.0744 |
| V | 399 | 199.5 | 0.554 | 5.40 | 0.814 | 0.931 | 0.874 | 0.054 |
| Al | 428 | 214 | .385 | 1.18 | 0.179 | 0.204 | 0.879 | 0.011 |
| La | 151 | 75.5 | 0.73 | 4.88 | 0.80 | 0.84 | 0.95 | 0.129 |
| $Nd_{2-x}Ce_xCuO_4$ | 409 | 204.5 | 0.897 | 22 | 3.7 | 3.79 | 0.976 | 0.215 |

*Both the McMillan and Allen-Dynes compositions are represented in Tables I and II.*

*Table Ia*

| Amorphous Superconductors | Reference | $\omega\log$(°K) | $E_o = \omega\log/2$ | $\lambda^{1/2}$ | $T_c$ | $2T_c/E_o$ |
|---|---|---|---|---|---|---|
| $Pb_{0.60}Tl_{0.40}$ | 24 | 50 | 25 | 1.175 | 5.90 | 0.472 |
| $Pb_{0.90}Bi_{0.10}$ | 11 | 50 | 25 | 1.288 | 7.65 | 0.612 |
| $Pb_{0.80}Bi_{0.20}$ | 11 | 46 | 23 | 1.371 | 7.95 | 0.691 |
| $Pb_{0.70}Bi_{0.30}$ | 11 | 47 | 23.5 | 1.418 | 8.45 | 0.719 |
| $Pb_{0.65}Bi_{0.35}$ | 11 | 45 | 22.5 | 1.459 | 8.95 | 0.796 |
| $In_{0.90}Tl_{0.10}$ | 11 | 63 | 31.5 | 0.894 | 3.28 | 0.208 |
| $In_{0.73}Tl_{0.27}$ | 11 | 55 | 27.5 | 0.964 | 3.36 | 0.244 |
| $In_{0.67}Tl_{0.33}$ | 11 | 57 | 28.5 | 0.949 | 3.26 | 0.229 |
| $In_{0.5}Tl_{0.5}$ | 11 | 53 | 26.5 | 0.911 | 2.52 | 0.190 |
| $In_{0.27}Tl_{0.73}$ | 11 | 42 | 21 | 1.044 | 3.64 | 0.346 |
| $\alpha\, Pb_{0.45}Bi_{0.55}$ | 11 | 29 | 14.5 | 1.609 | 7.00 | 0.965 |

*NOTE: Most of the references ($\theta_D$ and $T_c$) for these compounds can be found in the Handbook of Superconductivty and other properties are referred to throughout the paper.*

*Figure 2*

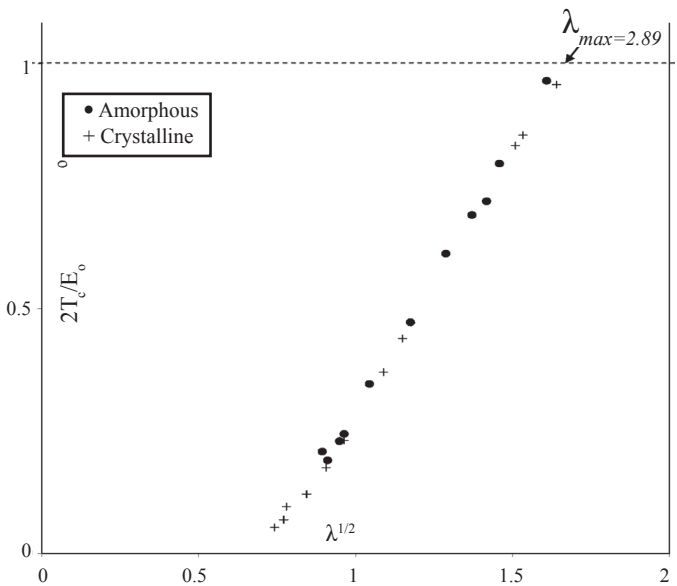

*Figure 2 illustrates the response of the dimensionless $2T_c/E_o$ against $\lambda^{1/2}$ for a variety of superconductors both crystalline and amorphous, wherein $\lambda$ is derived from our universal equation $T_c = E_o \exp(-2/\lambda)$ reflecting the efficiency of utilization of phonon energy in the formation of Cooper pairs.*

We further observe that at our predicted $\lambda_{max} = 2.89$ is within these relationship or that $2T_c/E_o = 1.0$ as indicated in Figure 2. Similar observations have been confirmed by several authors including M. Casas[10] et al whose views are arrived at by independent methodology but yield similar ratio and lead to a similar conclusion as well, i.e. that BCS-Bose crossover occurs at $\lambda > 2.89$ and is consistent with a similar view by G. Varelogiannis and L. Pietronero[11], where they take the view that a $\lambda \geq 3$ represents the beginning of the crossover region into Bose condensation region. Both of these references are in essential quantitative agreements with our view of the upper limit of BCS-electron phonon mediated superconductors at $\lambda = 2.89$.

## Further Elaboration of the Significance $2T_C/E_O$ and $\exp(-2/\lambda)$

We have previously indicated and we now illustrate the efficiency of utilization of available phonon energy, the components of which are ½ $\hbar\omega_D$ = zero point energy or in temperature units $E_o$. Whereas $\Delta_{c(o)}$ is the coherent condensation energy gap which also is equal to $2T_c$ i.e.
Eq. 7: $2T_c = \Delta_{c(o)}$

and for the efficiency of phonon utilization $\eta = 2T_c/E_o = 1.0$ for which the efficiency of utilization is 100% or on a normalized scale 1.0. Thus, a plot of $\lambda^{1/2}$ vs. $2T_c/E_o$ as shown in Figure 2 intercepts unity at a $\lambda^{1/2} = 1.70$ or $\lambda = 2.89$. Thus we conclude that the maximum possible $\lambda$ for electron-phonon mediated superconductors is 2.89. This is an interesting value since Allen-Dynes[9] in their amorphous materials observed a $\lambda = 2.59$ for $Pb_{0.45}Bi_{0.55}$, we therefore have experimental values measured by tunneling that approach closely the maximum value of 2.89 and are independent of the crystalline state of the superconductor which may be either crystalline or amorphous. Our equation for $T_c = E_o \exp(-2/\lambda)$ applies indifferently to both types. Interestingly $Tl_2Ba_2CuO_{6+3}$ whose $\theta_D = 363°K$ and whose $T_c = 90°K$ and $\lambda = 2.85$ and a $2T_c/E_o$ value slightly in excess of 0.99 i.e. it is approximately 100% efficient.

It is thus evident that one of the principal differences distinguishing Cuprates from other superconcuctors is that they operate at an efficiency of electron-phonon energy utilization approaching 100% as compared to other superconductors at $\leq 30\%$.

### Verification of Eq.5 with Amorphous Superconductors

We now go the Allen-Dynes[9] data in analogy with the McMillan plot we plot $\ln(\omega_{log}/2T_c)$ vs. $\lambda^{-1}$ (Allen- Dynes). For this data set we arrive at Figure 3 which is linear with intercepts $0,0$ for $\lambda^{-1}$ vs. $\ln(\omega_{log}/2T_c)$ and is represented by equation 6 below.
Eq. 6: $T_c = (\omega_{log})/2\exp[-2/\lambda]$

The Allen-Dynes[9] parameters are found in Table Ia.

We thus arrive at a universal equation whose exponential component is identical to that arrived at with the McMillan[7] data set and whose pre exponential has the same form i.e. a (characteristic Temperature)/2 which in the case of the McMillan data is $\theta_D/2$ and in the case of the Allen-Dynes data $(\omega_{log})/2$, where $\omega_{log}$ is in units of °K.

*Although Allen-Dynes[9] exponential prefactor is characterized as $\omega_{log}$ it is in fact the equivalent Debye Temperature of their amorphous materials.*

We have thus arrived at a universal equation for phonon mediated superconductivity where there are no adjustable variables. And, that represent either Crystalline or Amorphous Superconductors. We also note that $\omega_{log} \simeq 0.53$ ($\theta_D$ Crystalline) thus establishing a relationship between amorphous and Crystalline Superconductors with a universal equation of the same form. Thus, taking into account that the Allen-Dynes[9] values are the effective Debye temperature for their materials the Allen-Dynes[9] and McMillan[7] ana-

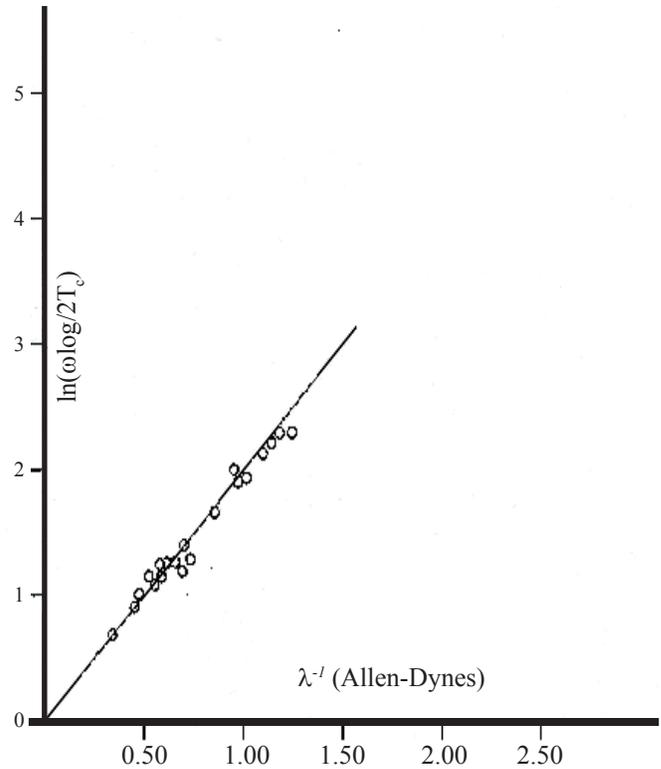

*Figure 3*

*Figure 3 illustrates the response of the ln of the dimensionless ratio $\omega_{log}/2T_c$ (wherein $\omega_{log}$ is in units of degrees kelvin) to the $\lambda_{(Allen-Dynes)}$ value at an approximate constant $\mu^* = 0.10$ for a variety of amorphous superconducting compounds.*

lytic solutions are identical. Thus, <u>unifying</u> the principles governing the superconductivity of crystalline and amorphous superconductors.

Thus, we find that our universal equation 6 for $T_c$ produces the same $\lambda$'s as does the McMillan equation for Crystalline <u>materials</u> and the Allen-Dynes equation for amorphous materials which heretofore was not possible without the unifying characteristics of equation 5, ie $T_c = \theta_D/2 \exp(-2/\lambda)$.

We illustrate the versatility of the basic equation and basic relationship of $\lambda^{-1}$ vs. $\ln(\theta_D/2T_c)$ for a wide variety of materials from amorphous to Crystalline to A-15 and cuprates as well as various elements with a range of $\lambda$ from 2.85 to 0.148 and $T_c$ from $325\mu°K$ to $133°K$ as shown in Figure 4 which also accommodates the electron doped cuprates as evidenced by $Nd_{1-x}Ce_xCuO_4$.

We now proceed to an essentially independent approach to further confirm and/or establish the universal expression for $T_c$. Cooper[13] in 1956 published a strong coupling expression which was illustrated and explained by Schrieffer[14] in 1976 wherein:

Eq. 8: $E_B \sim \hbar \omega_D \exp[-2/NoV]$

Where $E_B$ = pair binding energy. Schrieffer[14] further assumed that the pairing was restricted within an energy $\hbar \omega_D$ of the Fermi energy as in BCS theory. Further Schrieffer notes if $E_B$ is taken to be of the order of $k_B T_c$ we can then arrive at a modified expression i.e.

*Figure 4*

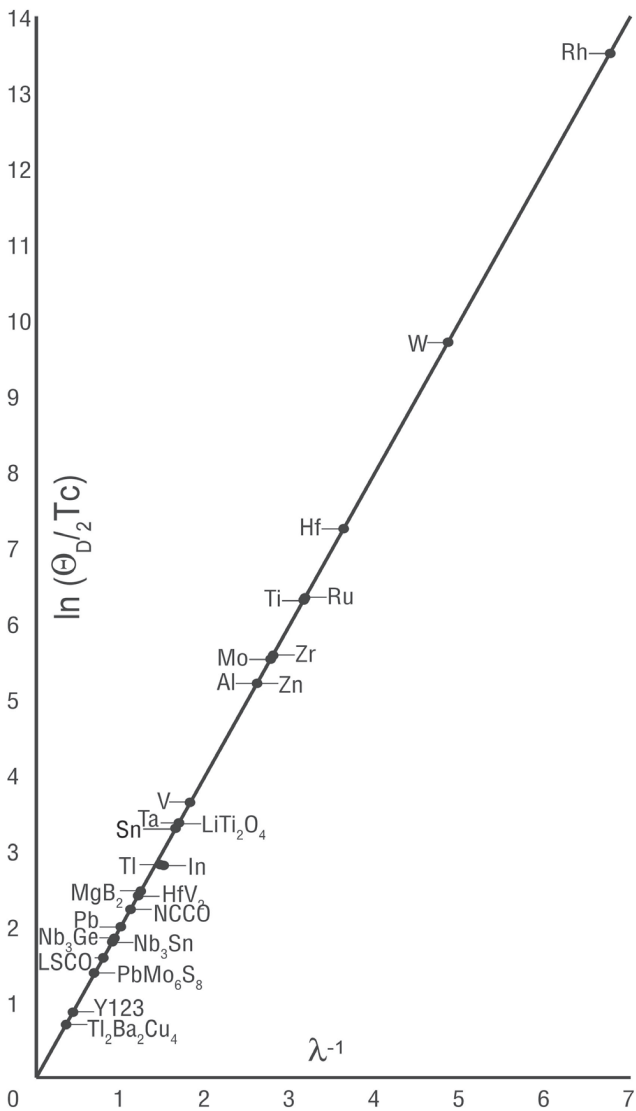

*Figure 4 illustrates a large variety of both amorphous and crystalline superconductors and cuprate superconductors as well as A-15 compounds and $PbMo_6S_8$ and others wherein $\lambda$ is derived from our universal equation $T_c=E_o \exp(-2/\lambda)$ and related to the dimensionless quantity $ln(\theta_D/2T_c)$.*

Eq.9: $T_c = \theta_D \exp[-2/NoV]$
Where NoV is equivalent to $\lambda - \mu^*$

Alternatively by the law of corresponding states we can also accurately establish the prefactor for Equation 9. If we equate NoV= $\lambda_{McMillan}$ and solve equation 9 for a number of $\lambda_{McMillan}$ values for various elements for $\lambda < 1.0$ and $> 0.35$. i.e. the range of validity of the McMillan equation, we find on average a value of $T_c$ that is a factor of 2 too high, that is, for equation 9 to accurately reflect $T_c$ for proper input $\lambda_{McMillan}$ values the prefactor necessarily becomes $\theta_D/2$ as in all of our previous cases. Thus from an independent semi-empirical route we again arrive at our universal equation for $T_c$. The generality of the prefactor $\theta_D/2$ and the exponential component of our Tc equation as $\exp(-2/\lambda)$ is thus strongly confirmed

It becomes interesting to note Schrieffer[14] proceeds to comment on Cooper's original equation, thusly "*While Coopers result was highly suggestive a major problem arose. If, as we discussed above a fraction $10^{-4}$ of the electrons is significantly involved in the condensation, the average spacing between these condensed electrons is roughly $10^6$ cm. therefore within the volume occupied by the bound state of a given pair the centers of approximately $(10^{-4}/10^{-6})^3 \sim 10^{-6}$ other pairs will be found on average. Then rather than a picture of a dilute gas of strongly bound pairs, quite the opposite picture is true. The pairs overlap so strongly in space that the mechanism of condensation would appear to be destroyed due to the numerous pair pair collisions interrupting the binding process of a given pair*".

Apparently such a situation is approached in strong coupling cuprates and one would expect that at strong coupling i.e. $\lambda \sim 2.5$ which we observed in cuprates that $T_c < T_p$ that is the pairing energy, $T_p$, is greater then the coherent condensation energy at $T_c$ thus accommodating some pair destruction. Whereas in superconductors with smaller $\lambda < 1$ generally $T_c \geq T_p$. We find that in a strong coupling cuprate such as $YBa_2Cu_3O_7$ that the pairing energy (Temperature) exceeds the coherent condensation temperature $T_c$ by $\sim 24°K$. i.e. $T_c = 92°K$ and $T_p \approx 116°K$, thus instead of the mechanism of condensation being destroyed as Schrieffer suggested that, in fact it is merely suppressed. i.e. as the coupling constant $\lambda$ increases so will the interval between the $T_p$ and $T_c$ generally increase as well. This point will be illustrated subsequently.

The likely possibility for the suppression of $T_c$ of strongly coupled superconductors is that "phase coherences" may require a lower temperature and when achieved, the coherent condensation can occur.

### The $\Delta_o/T_c$ Problem

Both, in BCS Theory and in the literature of Superconductivity $\Delta_o$ is referred to as the coherent condensation energy gap and in BCS its value is referenced to $T = 0°K$ and, in experimental literature $\Delta_o$ is referred to $0°K$ as well. Also commonly used is the ratio $\Delta_o/T_c$.

In Appendix 1 we illustrate experimental values of $\Delta_o$ and of $T_c$ and their ratio for many elements and compounds and note that their ratio varies from ~1.7 to 2.3. Here, is becomes necessary to redefine $\Delta_o$ as $\Delta_{E(o)}$ the experimentally measured gap energy wherein $\Delta_{E(o)} = \Delta_{c(o)} \pm \Delta_x$ where $\Delta_{c(o)}$ is the coherent condensation energy gap and $\Delta_x$ is an additional energy term reflecting whether the superconductor is weak or strong coupling. For strong coupling superconductors $\Delta_x$ is positive indicating the detection of pairs above $T_c$ the coherent condensation temperature i.e. $T_p$(pairing temperature) exceeds $T_c$ (critical temperature or the coherent condensation temperature) and this occurs only with strong coupling superconductors such as Hg or Pb. Whereas for weak superconductors such as Al. From experimental measurements it transpires that $T_p < T_c$. In other words for strong superconductors with $\lambda$ values greater than 0.77 the coherent condensation is suppressed whereas for weak superconductor with $\lambda < 0.77$ coherent condensation is enhanced. The $\Delta_{c(o)}$ value can be extracted from the $\Delta_{E(o)}$ value by a simple method. We redefine the ratio $\Delta_{E(o)}/T_c$ as $\Delta_{E(o)}/2T_c$ which essentially normalizes the ratio to unity, i.e. strong superconductors have a ratio >1.0 and weak superconductors a ratio of $\Delta_{E(o)}/2T_c < 1.0$. Thus the departure from unity yields the necessary correction term to establish $\Delta_{c(o)}$ from $\Delta_{E(o)}$.

For example Hg has a $T_c = 4.15°K$ and a $\Delta_{E(o)} = 0.824$mev or $9.56°K$, therefore $\Delta_{E(o)}/2T_c = 9.25/8.3 = 1.152$ thus $\Delta_{c(o)} = \Delta_{E(o)}/1.152 = 9.56°K/1.152 = 8.30°K$ and $\Delta_{c(o)}/T_c = 8.30/4.15 = 2.00$ that is for Hg $\Delta_{c(o)} = 2k_B T_c$ as we found in our earlier analysis.

For Pb its $T_c = 7.2°K$ and $\Delta_{E(o)} = 1.33$mev or $15.43°K$ therefore $\Delta_{E(o)}/2T_c = 15.43/14.4 = 1.071$ thus $\Delta_{c(o)} = \Delta_{E(o)/1.071} = 1.24$mev or $14.4°K$. Again we have $\Delta_{c(o)}/T_c = 2$ and $\Delta_{c(o)} Pb = 2k_B T_c$ this simple form of correction is valid for all phonon mediated superconductors including cuprates.

We note in particular that for $f\lambda > 0.77$ that the departure of $\Delta_{c(o)}/2T_c$ increases almost linearly and monotonically with increasing $\lambda$ up through and including the cuprates and Hg and Pb. Thus indicating no mechanism change in this parameter as it passes from strong coupling elements to the cuprates. Thus, reinforcing our conclusion that the cuprates are electron phonon mediated.

### Relationship of $\Delta_{E(O)}/2T_C$ VS. $\lambda^{1/2}$ and its Significance to Cuprate Superconductors

Figure 5 illustrates the relationship between $\Delta_{E(o)}/2T_c$ vs. $\lambda$ which is linear throughout its complete range from weak to very strong superconductors i.e. from Al to $Tl_2Ba_2CuO_6$ where $\lambda$ values are 0.385 and 3.85 respectively and include diverse superconductors such as elements, A-15 compounds ($Nb_3Sn$ and $Nb_3Ge$) as well as $MgB_2$ and $PbMo_6S_8$.

The lack of any discontinuity or slope change in progressing from normal electron phonon mediated superconductors such as Al, Nb, Sn, Pb, and $MgB_2$ to the cuprates including LSCO, Y123 and $Tl_2Ba_2CuO_6$ provides compelling evidence that cuprates are:
1. electron phonon mediated,

2. that no change in energy scales for the cuprates is required to explain their superconductivity i.e. exitonic or other mechanisms,

3. that the psuedogap associated with all cuprates plays no role in their superconductivity.

W. S. Lee et al have reported the discovery of a coherent condensation energy gap in slightly underdoped BSCCO with a $T_c = 92°K$. This Gap measured in the nodal direction exhibits a cannonical BCS-like temperature dependence accompanied by a Bugoluibov quasi-particle signature in contrast to the gap measured in the anti-nodal direction near the Cu-O bond direction (the anti-nodal region) which has been typically measured in this anti-nodal direction in earlier experiments by many others. The anti-nodal gap does not show a significant temperature dependence and has a significantly larger value of the gap energy. This two-gap scenario provides direct and compelling evidence of a coherent condensation energy gap of 16mev ± 1 which clearly indicates that $\Delta_{c(o)} = 2k_B T_c$ since 92*K is equivalent to 15.86 mev in good accord with our values of $\Delta_{c(o)} = 2k_B T_c$. The referenced two-gap ARPES experiment is a clear validation of the non role of the psuedo-gap in the superconductivity of cuprates.

The above evidence is even more compelling when one considers that $\Delta_{E(o)}/2T_c$ and $\lambda^{1/2}$ are values derived from essentially independent experimental data bases and again confirm the internal self-consisitency for all of our universal expressions for superconductivity.

## Cuprate Superconductors- Calculation of Hole Concentration, P, Required to Achieve Maximum $T_c$ FOR A CUPRATE SUPERCONDUCTOR AND P vs. $T_C$

We chose the well studied $La_{2-x}Sr_xCuO_4$ system to demonstrate the methodology to arrive at the hole concentration required for the maximum $T_c$ of the $T_c$ vs. hole doping function.

Our starting point is that experimentally this maximum $T_c \approx 38°K$. We now need to find the change in chemical potential from $La_2CuO_4$ to produce a $T_c$ of 38°K. The usual definition of chemical potential is used in this calculation which is related to the Mulliken absolute electronegativity as follows.

Eq. 10: $\mu_{Mulliken} = -\chi_{Mulliken} = -\dfrac{IP + EA}{2} = \left[\dfrac{\delta E[N]}{\delta N}\right]_{N=N_o}$

Thus, the Mulliken chemical potential is a finite difference approximation of the electronic energy with respect to the number of electrons. Where $\mu_{Mulliken}$ is the chemical potential and $\chi_{Mulliken}$ is the electronegativity. We use the electronegativity values since they differ in sign only. Thus we need only know the electronegativity values of the constituent atoms of compounds of interest, these are 3.02ev, 2.87ev, 4.48ev and 7.54ev for La, Sr, Cu and O respectively.

*Figure 5*

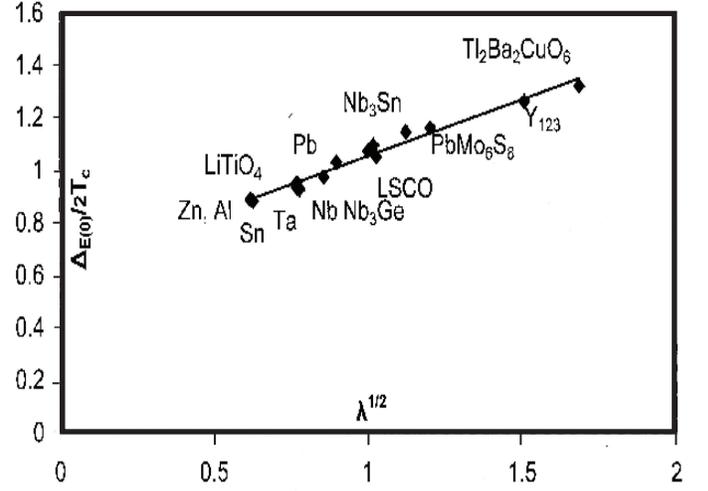

*Figure 5 illustrates the relationship of $\Delta_{E(o)}/2T_c$ vs. $\lambda^{1/2}$ for a wide variety of superconductors from weak to strong coupling.*

We need only to know the $\chi$ values of two end member compounds which are $La_2CuO_4$ and $LaSrCuO_4$ the $\chi$ values calculated for each compound are based on the principal of electronegativity equalization or chemical potential equalization and are effectively the geometric mean $\chi$ value for each compound. For $La_2CuO_4$, $\chi = (3.02^2 \times 4.48 \times 7.54^4)^{1/n}$ where n=7 the number of atoms in the compound thus, $\chi(La_2CuO_4) = 5.3894$ ev. For $LaSrCuO_4$, $\chi = (3.02 \times 2.87 \times 4.48 \times 7.54^4)^{1/7}$ = 5.3503 and $\Delta\chi = 0.0391$ev or 39.1mev. From $\Delta_{c(o)} = 2T_c$ we then have 38°K = $T_c$ and $\Delta_{c(o)} = 6.55$mev thus the hole concentration which is x in $La_{2-x}Sr_xCuO_4$ required to achieve: $T_c$max = 38°K is

Eq. 11: $X_{Tc}$max = $\dfrac{6.55\ mev}{39.1\ mev}$ 0.1675

This calculated hole or Sr concentration is in excellent agreement with experimental values from the literature. Since we know the required parameters and can calculate the $\Delta\mu$ for any Sr concentration we can generate the entire dome shaped curve by solving Eq 12.

Eq. 12: $T_c = 38°K \left[1 - \dfrac{|\Delta\mu_{T_c\max} - \Delta\mu_x|}{\Delta\mu_{T_c\max}}\right]$

whose solution is shown in Figure 6 along with comparative experimental $T_c$'s from the literature. Equation 12 via Figure 6 illustrates excellent agreement with the experimental values. There are two noteworthy points and these are:

1. The doping curve for $La_{2-x}Sr_xCuO_4$ results in $T_c=0$ at a hole concentration=0.32 and at $2\Delta_{c(o)}$ as well. Indicating clearly that $\Delta\mu$ values in excess of $\Delta\mu_{peak}$ required for a maximum $T_c$ result in increasing pair destruction up to $2\Delta\mu_{peak}$ at which $T_c$ returns to zero and all pairs existing at $\Delta_{c(o)}$ are now broken.

2. The value of $\Delta\mu_{peak}$ calculated is exactly equal to $\Delta_{c(o)}$ experimental i.e. 6.55mev which we obtain from $\Delta_{E(o)}$ values of $7^{17}$ and $8mev^{18}$ averaged or 7.5mev and where we found that $\Delta_{E(o)} 7.5mev/2T_c=7.5/6.55=1.145$ thus the $\Delta_{c(o)}$ value of 6.55mev for $La_{2-x}Sr_xCuO_4$. Thus validating our methodology for calculating the $\Delta\mu$. Again reinforcing that $La_{2-x}Sr_xCuO_4$ is an electron phonon mediated superconductor.

The data source for $T_c$ vs. Sr concentration is provided by D. G. Radaelli,[16a] Y. Wang et al,[16b] and T. Tsukada et al.[16c]

*Figure 6*

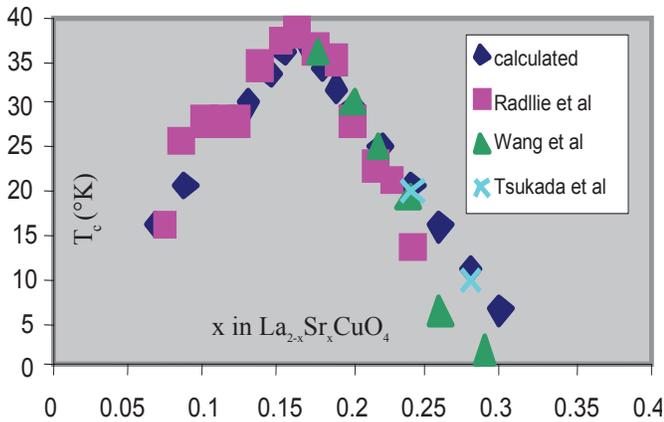

*Figure 6 (color) represents the $T_c$ response to Strontium substitution for Lanthanum in LaSrCuO for both equation 12 and experimental data from several sources. Our calculated value of x= 0.1675 for T=38K and teh experimental data are in excellent agreement with those of Randelli et al[13], Y. Wang et al[14], and I. Tsukada et al[15].*

It is certain that x= 0.1675 is a direct consequence of the coherent energy gap, and furthermore at x= 0.32 where $T_c$ goes to zero in the overdoped region represents $2\Delta_{c(o)}$, i.e. one $\Delta_{c(o)}$ to peak superconductivity and $2\Delta_{c(o)}$ to zero superconductivity i.e. $T_c=0$. The shift in chemical potential $\Delta\mu_{coherent}=\Delta_{c(o)}$ at any Sr content along the doping curve. And, even though a gap exists for $x \leq 0.06$ no superconductivity is observed in this interval of x primarily due to values of resistivity that are too high and superconductive onset only occurs when resistivity reaches some lower critical value.

Having calculated the hole concentration for $La_{2-x}Sr_xCuO_4$ or the Sr concentration in a hole doped superconductor, we can possibly gain some insight into the fundamental reasons for hole superconductivity being more effective than the electron doped superconductivity in the cuprate materials, accordingly, we now calculate by similar methodology. The electron concentration in $Nd_{2-x}Ce_xCuO_4$, a compound of identical structure but doped for electron rather than hole conductivity. The essential elements are the $\chi$ values of Nd and Ce and the two end-point compositions i.e. $Nd_2CuO_4$ and $NdCeCuO_4$ where respective $\chi$ values are 5.4547 *mev* and 5.4791 *mev* for a $\Delta\mu$ of 24.4 *mev* where $\chi Nd = 3.15$ *ev* and $\chi Ce = 3.25$ ev with the Cu and oxygen $\chi$ values as before in the LSCO calculation. Thus the electron concentration at peak $T_c$ of 22°K is $2T_c$ or 44°K which translates to 3.79 *mev* therefore, electron concentration at peak $T_c$ is equal to $\dfrac{3.79\ mev}{24.4\ mev} = 0.155$ which is verified by reference 33 as well. Thus, for approximately equal hole and electron concentrations i.e. for LSCO p = 0.1675 and for NCCO Ce concentration is equal to 0.155 and the $\theta_D$'s are 368°K and 409°K respectively.

The values of Sr or Ce concentration at peak $T_c$ values for $La_{2-x}Sr_xCuO_4$ and $Nd_{2-x}Ce_xCuO_4$ are reported by B. Lake et al[19] and D. K. Sunco et al[20] and are 0.16 and 0.15 respectively.

We also find that for these two compositions $\eta = 2T_c/E_o$ is 0.107 and 0.206 for NCCO and LSCO respectively at their optimal $T_c$ values. Thus, when accounting for the $\theta_D$ difference, the residual efficiency factors completely account for their $T_c$ differences. The result of removing the $\theta_D$ differ-ences and multiplying by the residual normalized η's which leaves us with $(\eta_{LSCO}/\eta_{NCCO})$ at equal $\theta_D$ = 1.76 or $T_{c\,LSCO}$ = 1.76 $T_{c\,NCCO}$ = 1.76x22=38.65 within 1.7% of the true value of optimal LSCO as deduced by relative efficiency of available phonon energy utilization.  We now may understand the primary reason that hole doped superconductivity is more effective than equivalent eletron doped superconductivity in cuprates.  The theoretical reasons for this efficiency difference between electron and hole doping remain unclear but may be related to the fact that the hole velocity is less than the electron velocity.

## Comparison of Current Finding to the Bcs View

Having developed from experimental data fundamental changes in the equations of superconductivity we have as a result two major changes in the expression of the exponential component describing $T_c$. From a value of $T_c \propto \exp(-1/\lambda)$ to $T_c \propto \exp(-2/\lambda)$ and also to the prefactor for the $T_c$ equation i.e. from BCS value of $T_c \propto 1.13\theta_D$ to our expression of $T_c \propto E_o$ or $\theta_D/2$ resulting in our final expression of *$T_c = E_o \exp(-2/\lambda)$ has profound implications respecting its ability to accommodate essentially the entire range of λ from <0.20 to greater than 2.8 without significantly changing the values for the McMillan $\lambda_{Mc}$ or those of Allen-Dynes which are tied to experimental data derived from tunneling in experimental electron-phonon mediated superconductors.*

We now proceed to demonstrate that the major cuprates for which exist reliable values of the experimental energy gap $\Delta_{E(o)}$, i.e. $La_{1.85}Sr_{0.15}CuO_4$ and $YBa_2Cu_3O_{6.95}$ as well as $Tl_2Ba_2CuO_6$, $Nd_{2-x}Ce_xCuO_4$, and other cuprates for which additional evidence, (including INS-Inelastic Neutron Scattering[20] and IXS-Inelastic X-ray Scattering[21]) and these will be subsequently discussed.

We revisit the following superconductors: $La_{1.85}Sr_{0.15}CuO_4$, $YBa_2Cu_3O_{6.95}$, and $Tl_2Ba_2CuO_6$ with respect to the elements of our universal equation for electron-phonon mediated superconductivity. The following Table II represents their accommodation within our universal analytic expression for superconductivity.

And for these latter superconductors there also exists compelling evidence that pairing within them is exclusively electron-phonon mediated.

The cuprates as can be seen by being referenced to the correct energy scale that is $E_o$, has a completely linear relationship regarding the ratio of $2T_c/E_o$ which we have previously defined as the fractional utilization of the available lattice energy ($E_o$) utilized in electron phonon coupling. The efficient hole doped cuprates are accommodated in this relationship and for $T_c$'s between 90°K-135°K the efficiency ranges from $\cong$ 80% to 99% respectively. The largest efficiencies we observed with strong coupling cuprate superconductors i.e.~80 to 99%; whereas elements or compounds that are non cuprates or electron doped cuprate range from 21% to 35%. The hole doped cuprates are approximately 2 times as efficient in achieving effective lattice interaction resulting in pairing than the most efficient elements or non cuprate compounds for example; $Nb_3Sn$ is ~31% efficient in comparison.

With four of the listed superconductive cuprates there exists independent evidence that the electron phonon interaction is the principal, if not exclusive, means for pair formation. Additionally, the phonon role for several cuprates is clearly demonstrated by inelastic X-ray scattering by T. Fukuda et al (Cond-Mat.-0306190)[16] wherein the authors concluded that these observations show direct evidence of full correlation of the LO phonon mode with superconductivity leading to the conclusion that phonon modes play a crucial role in the superconductivity of LSCO and probably other cuprates as well.

Giag-Meng Zhao[21] has demonstrated that the second derivative of the tunneling current $d^2I/dV^2$ show clear dip and peak features due to strong coupling to the Bosinic modes indicating electron pairing. The energy positions of which those of the phonon density of states obtained by inelastic neutron scattering for both $YBa_2Cu_3O_7$ and $Bi_2Sr_2CaCu_2O_8$ indicating clearly that high temperature superconductivity should arise primarily if not exclusively from strong coupling to phonon modes.

Also tunneling spectroscopy of $Tl_2Ba_2CuO_{6+3}$ by L. Ozyuzer[22] and D. Mazor[22a] yields clear evidence of a measured gap value of 20.5mev ±1 with a minimum $T_c$ of 86°K. And this evidence coupled with Kulkarni et al $\theta_D$ value of 363°K[23] yielding a minimum λ of 2.68 and a $2T_c/E_o$ value equals 0.948 accommodates well to our universal analytic equations for electron-phonon mediated systems.

Although the above provides excellent evidence of the major, if not dominant, role of phonons in cuprates. We further elaborate this point by demonstrating the universal connection between what appear as disparate superconductors by calculating the $T_c$ of $YBa_2Cu_3O_7$ from either the parameters associated with Aluminum, ($\theta_D$=428°K, $T_c$= 1.18°K and λ=0.385) and or $Pb_{.45}Bi_{0.85}$ $\theta_D$= 29°K, $T_c$=7.0°K, and λ=2.59 ($\theta_{D(Y123)}$=437°K $T_c$=92°K, and λ=2.312).

Eq. 10: $T_c$ of Y123 = $\left(\dfrac{\theta_{D(Y123)}}{\theta_{D(Al)}}\right)\left(\dfrac{\exp(-2/2.315)}{\exp(-2/0.385)}\right)(T_{c\,of\,Al})$ = 91.37°K

and using a $Pb_{0.45}Bi_{0.55}$ as the base we have:

$T_{c(Y123)} = \left(\dfrac{\theta_{D(Y123)}}{\theta_{D(Pb.45Bi/.55)}}\right)\left(\dfrac{\exp(-2/2.315)}{\exp(-2/2.59)}\right)(T_{c\,(Pb.45Bi/.55)})$ = 96.09

The calculation of the $T_c$ of Y123 from the superconductor properties of Tungston which are $T_c$=0.012°K and $\theta_D$= 390°K, and using W as the base we have:

$T_{c(Y123)} = \left(\dfrac{\theta_{D(Y123)}}{\theta_{D(W)}}\right)\left(\dfrac{\exp(-2/2.315)}{\exp(-2/0.00002068)}\right)$ x 0.012 = 91.85°K

In both cases the $T_c$'s of Y123 calculated from the properties of three extreme superconductors within 4% of the accurate value of 92°K for Y123 yielding clear indication of the continuity of mechanism throughout.

*Most importantly, given, the derivation of our self consistent and universal superconductivity equations emanating from a Debye lattice vibrational mode dominated genesis, there can be no doubt that cuprates are electron phonon mediated superconductors since they were accommodated within our expression with no difficulty or unnecessary adjustments and are accommodated easily into $2T_c/E_o$ as well.*

The expectation of the zero point energy which is identical to $E_o$ as an integral part of superconductive electron pairing was first anticipated by K.S. Pitzer[25]. In 1956 and precedes Cooper's 1956 paper. Curiously Pitzer's publication entitled "The Fundamental Theory of Superconductivity" was not referenced by BCS in their seminal 1957 paper. The Pitzer paper is the only one which conveys clearly the idea that <u>pairing and zero point energy are both important aspects of superconductivity.</u>

### Examination of the Properties Of Major Cuprate Compositions Within the Context of Our Self Consistent Equations for Electron-phonon Mediated Superconductivity

Table II and IIa illustrate the various significant parameters for the afore mentioned cuprates. We have selected those for which we believe reliable values of $T_c$, $\theta_D$ and in four cases, reliable values of $\Delta_{c(o)}$ are available. There are several notable features for superconductors for which we have reliable values of $\Delta_{c(o)}$ i.e. $YBa_2Ca_3O_7$, $Tl_2Ba_2CuO_6$, $La_{1.85}Sr_{.15}CuO_4$, and $Nd_{2-x}Ce_xCuO_4$: they are clearly consistent with $\Delta_{c(o)}$=$2T_c$ from experimental data. This suggests that the $\Delta_{E(o)}$ values measured by tunneling are too high for many cuprates and that the pseudogap which we believe has no relevant function vis a vis cuprate fundamental $\Delta_{E(o)}$ values but only interferes in the measurement and determination of $\Delta_{c(o)}$ values. On our last two superconductors i.e. $Tl_2Ba_2CaCu_2O_8$ and $Tl_2Ba_2Ca_2Cu_3O_{10}$ have $\theta_D$ values relative to their $T_c$ values that give values of λ of ~3.05 and 3.3 respectively we suggest their Debye temperatures are too low by ~5-10% which is not unusual at all in measurement of Debye temperatures. The rest of the cuprates fall neatly within the context of maximum λ=2.89 in fact the vast majority of cuprates do. The Bismuth cuprates also have slightly low Debye temperatures as well but if one observes Ledbetter's[26] data on the cuprate values of $\theta_D$ we note that the Bismuth compounds are not well behaved with respect to $\theta_D$ as the number of Cu layers increases and accordingly are apparently difficult to deal with in the experimental determination of their Debye temperatures. We note that $Bi_2Sr_2CaCu_2O_8$ has a $\theta_D$~288°K and a $T_c$~ 86°, one whose λ would be 3.87. For 86°K to fit our expres-sion where max $\lambda=2.89$, the required $\theta_D$ would be equal to 344°K a $\theta_D$ ~19% higher which we believe much more likely. Recall that $Tl_2Ba_2CuO_6$ has a $\theta_D$ ~363°K. The two compounds should be similar in that their trivalent radii are 0.95 for Tl and 0.96 for Bi. Ledbetter[26] also notes all of the superconductors were well behaved in that $T_c$ increased linearly with $\theta_D$ increase for all families with the exception of the Bismuth cuprates for which the effect had considerable scatter as found by Daminec[19] probably indicating more scatter and uncertainty in the measurements of $\theta_D$ since $T_c$ can be accurately measured. In general, the cuprates are well represented by our self consistent analytic equation and are without question electron-phonon mediated. Table IIb illustrates the calculated $\Delta_{c(o)}$ values for the tested cuprates as well as determined from $\Delta_{c(o)}=2k_BT_c$.

With respect to the cuprates listed in Tables II and IIa the references for $\Delta_o$ values, and $T_c$'s are largely from "Handbook of Superconductivity ed. by C.P. Poole, academic press (2000) 447. For $\theta_D$ values of $Tl_2$ and Tl series the reference is A.D. Kulkarni et al, Phys Rev B **43** 5451 (1991). For the Hg cuprates: H. Ledbetter, Phys. Rev. C 235-240 1325 (1994) for $La_{2-x}Sr_xCuO_4$ "Handbook of Superconductivity" ed. C.P. Poole Academic Press 557 and T. Kato et al Phys. Rev. C 392 231 (2003) and in Y123 and LSCO L.M. Lei and H. Ledbetter NIST Publication. PB92112333 (cap o oxides and oxide superconductor Elastic and Related Properties) U.S. Dept of Commerce (1991)(NIST 3980) F.W. DeWette Phys. Rev. **42** 6707 (1990) for Y123 $\theta_D$.

*Table II*

| Superconductor | $T_c$ °K | $\theta_D$ °K | $\lambda_{eq.4}$ |
|---|---|---|---|
| $La_{1.85}Sr_{.15}CuO_4$ | 38 | 365 | 1.275 |
| $YBa_2Cu_3O_7$ | 92 | 447 | 2.253 |
| $Tl_2Ba_2CuO_6$ | 90 | 363 | 2.851 |
| $Tl_2Ba_2CaCu_2O_8$ | 110 | 424 | 3.048 |
| $Tl_2Ba_2Ca_2Cu_3O_{10}$ | 125 | 458 | 3.303 |
| $TlBa_2CaCu_2O_7$ | 80 | 508 | 1.731 |
| $TlBa_2Ca_2Cu_3O_9$ | 120 | 536 | 2.489 |
| $TlBa_2Ca_3Cu_4O_{11}$ | 114 | 550 | 2.27 |
| $HgBa_2CuO_{4.1}$ | 94 | 440 | 2.352 |
| $HgBa_2CaCu_2O_{6.22}$ | 126 | 530 | 2.69 |
| $HgBa_2Ca_2Cu_3O_{8.44}$ | 133 | 577 | 2.583 |

*Table IIa*

| Superconductors | $\Delta_{E(o)}$ mev calc | $\Delta_{E(o)}$ mev exp | $\Delta_{c(o)}$ mev |
|---|---|---|---|
| $La_{1.85}Sr_{.15}Cu_4$ | 7.51 | 7.5 | 6.55 |
| $YBa_2Cu_3O_7$ | 19.93 | 20.0 | 15.86 |
| $Tl_2Ba_2CuO_6$ | 19.99 | 22.0 | 15.52 |
| $Tl_2Ba_2CaCu_2O_8$ | 24.5 | 30.0 | 18.96 |
| $Tl_2Ba_2Ca_2Cu_3O_{10}$ | 28.1 | - | 21.55 |
| $TlBa_2CaCuO_7$ | 16.75 | - | 13.79 |
| $TlBa_2Ca_2Cu_3O_9$ | 26.3 | - | 20.69 |
| $TlBa_2Ca_3Cu_4O_{11}$ | 24.7 | - | 19.65 |
| $HgBa_2CaO_{4.1}$ | 20.49 | 24 | 16.72 |
| $HgBa_2CaCu_2O_{6.11}$ | 27.85 | 50 | 21.72 |
| $HgBa_2CaCu_3O_{8.44}$ | 29.30 | 75 | 22.93 |

Further publication of a pnictide composition of $La O_{0.93}F_{0.07}FeAs$ by T. Sato[28] et al, where the d band coherent condensation gap is noted at 4.1 mev, which is completely consistent with it's $T_c$ of 24°K and the quantity $\Delta_{c(o)}=2k_BT_c$ equal to 4.13 mev in complete agreement with our own universal equations. Thus again finding the pnictide to be normal elevation phonon mediated superconductors.

**Conclusions**

The finding from empirical data of a self-consistent set of relationships describing superconductivity wherein $\theta_D/2$ is the prefactor for all $T_c$ equations and $\hbar\omega_D$ is the prefactor for the expression of $\Delta_{c(o)}$ which is consistent with the zero point energy being the source energy adjacent to $E_f$ i.e. the source energy for the formation of Cooper pairs via electron-phonon interaction and virtual exchange of phonons which we identify as $E_o = 1/2\ \hbar\omega_D$ or $\theta_D/2$ depending upon the units required for a particular solution and is the residual vibrational energy at T = 0°K. Thus our equations are firmly rooted in electron-phonon mediation of superconductivity. The net result is a seamless flow from one superconductor to any other including crystalline to amorphous as witnessed by our ability to calculate or inter-relate the superconductivity parameters from any superconductor to that of any other. and consistent with this observation is the reinforcement and constancy of the expression $\Delta_{c(o)}=2k_BT_c$ which is confirmed by its theoretical verification by X.H. Zheng.[9a]

Through these relationships we find other significant findings such as the maximum possible λ for electron-phonon mediated superconductivity is equal to λ = 2.89. We also find that the cuprates are not only easily accomodated by our expression for phonon mediated superconductivity which is reinforced by every family of cuprates i.e. based on Tl and Hg families of cuprates are found to scale the $T_c$ values with the Debye temperature as well i.e. in each case an increase in $θ_D$ results in corresponding increases in $T_c$ again confirming the electron-phonon mediated superconductivity. We also confirm η the efficiency of utilization of available phonon energy in Cooper pair formation as the primary parameter accounting for the higher $T_c$'s of hole doped vs. electron doped cuprate superconductors of the same structure i.e. LSCO and NCCO wherein the electron or hole doping levels are shown to be equivalent i.e. 0.16 + -0.01.

The analytic equations we derive from experimental superconductivity data retain the essential paradyme shifting element of "Cooper pairing" while accommodating to the proper energy scale for electron-phonon interaction i.e. the zero point energy which results in the universality of our equations and accommodation of cuprates as electron-phonon mediated superconductors thus, resulting in a modified BCS mechanism of superconductivity.

Appendix 1

| Compound | $\Delta_o$ (mev) | $T_c$ (°K) | $\Delta_o$ (ln °K) | $\Delta_o/T_c = \Delta_{E(o)}/2T_c$ |
|---|---|---|---|---|
| Al | 0.179 | 1.16 | 2.07 | 1.784 |
| Ca | 0.072 | .42 | 0.83 | 1.976 |
| Zn | 0.13 | 0.85 | 1.51 | 1.776 |
| Ga | 0.169 | 1.06 | 1.96 | 1.849 |
| Sn | 0.593 | 3.72 | 7.29 | 1.96 |
| In | 0.541 | 3.41 | 6.28 | 1.841 |
| Tl | 0.369 | 2.38 | 4.28 | 1.798 |
| Pb | 1.33 | 7.14 | 15.44 | 2.162 |
| Hg | 0.824 | 4.16 | 9.59 | 2.305 |
| $Tl_{0.9}Bi_{0.1}$ | 0.355 | 2.3 | 9.12 | 1.791 |
| $Pb_{0.4}Tl_{0.6}$ | 0.805 | 4.6 | 9.34 | 2.03 |
| $Pb_{0.6}Tl_{0.4}$ | 1.08 | 5.9 | 13.17 | 2.23 |
| $Pb_{0.8}Tl_{0.2}$ | 1.28 | 6.8 | 14.84 | 2.18 |
| Re | 0.263 | 1.7 | 3.07 | 1.805 |
| Ta | 0.72 | 4.49 | 8.35 | 1.859 |
| V | 0.814 | 5.40 | 9.44 | 1.748 |
| Nb | 1.55 | 9.22 | 17.98 | 1.95 |
| $Nb_3Sn$ | 3.30 | 18.3 | 39.32 | 2.148 |
| $Nb_3Al$ | 3.04 | 16.4 | 33.26 | 2.028 |
| $Nb_3Ge$ | 4.36 | 23 | 48.25 | 2.097 |
| $UPt_3$ | 0.075 | 0.44 | 0.87 | 1.977 |
| La | 0.8 | 4.88 | 9.28 | 1.901 |
| $LiTi_2O_7$ | 1.8 | 11.0 | 20.88 | 1.898 |
| $Nd_{1-x}Ce_xCuO_4$ | 3.7 | 22.0 | 42.4 | 1.927 |
| $La_{1.85}Sr_{0.15}CuO_4$ | 7.0 | 36 | 81.2 | 2.255 |
| $YBa_2Cu_3O_7$ | 20 | 92 | 232 | 2.52 |
| $Ba_{0.6}K_{0.4}BiO_3$ | 4.5 | 28 | 53.8 | 1.92 |
| $Nb_{0.8}Zn_{0.2}$ | 1.94 | 11.0 | 22.5 | 2.045 |

NOTE: The DeWette calculation for $\theta_D=465.5$°K vs. Ledbetter's $\theta_D$ for Y123= 437°K. The calculated values can be used in the comparison of cuprates against most of the experimentally measured $\theta_D$'s, where the $T_c$ are comparable.

NOTE 2: $\Delta_{E(o)}/T_c$ is the experimental value of coherent condensation gap.

Appendix 2

Recently, a new class of high temperature superconductors ahve been discovered and range in $T_c$ from a few °K to ~55°K and can be electron or hole doped similiarly to the cuprates and are 2 dimensional as well. A prototypical compound is LaFeAsO which when doped with Flouride such that the resultant compound has the composition of $LaFeAsO_{0.90}F_{0.10}$ [27] with an onset $T_c$ of ~28°K. From specific heat data the $\theta_D$ of this compound has been determined to be 315.7°K we then can calculate it's λ from our equation $\lambda = -2/\ln(E_{(o)}/T_c)$ and find that λ =1.156. Thus, $LaFeAsO_{0.90}F_{0.10}$ fits perfectly figure 4 as well as conforming to $\Delta_{c(o)}= 2k_B T_c$ and is equal to 4.83 mev. Consequently it is completely consistent with the electron phonon mediation developed by our universal equations.

A further verification of the above is the calculation of the $T_c$ of Y123 from the parameters of LaFeAsO$_{0.90}$F$_{0.10}$ as follows;

$$T_c(Y123) = \left(\frac{\theta_{D(Y123)}}{\theta_{D(Pnictide)}}\right)\left(\frac{0.421}{0.1775}\right) \times 28°K = 1.384 \times 2.372 \times 28 = 91.91°K$$

where $0.421 = \exp(-2/\lambda_{123})$ and $0.1775 = -2/\lambda_{Pnictide}$.

Accordingly, we conclude that the Pnictides are exclusively electron-phonon mediated and require no universal source energy to accomodate their superconductivity.

Further publication of a Pnictide composition of LaO$_{0.93}$F$_{0.07}$FeAs by T. Sato et al[28] where the dband coherent condensation gap is noted to beat 4.1mev which is completely consistent with it's $T_c$ of 24°K and the quantity $\Delta_{c(o)}$ is equal to $2k_B T_c$ equal to 4.13mev in complete agreement with our own universal equations. Thus again finding the Pnictide to be a normal electron-phonon mediated superconductor.

*end*